# Determining optical properties of skin ex-vivo: An inverse solution

PARIKSHAT SIRPAL*

*Polytechnique Montreal, Montreal, Canada*

*parikshat.sirpal@polymtl.ca

ABSTRACT

A rapid and accurate numerical solution to the transport equation (i.e., inverse adding doubling) of diffuse reflectance has been utilized here for the calculation of optical properties of the Yucatan miniature pig skin as a human surrogate skin. This method is valid for a wide range of optical properties and can be adapted to a variety of probe geometries, provided calibration measurements are made. Inverse adding doubling (IAD), based on the forward solution to the transport equation is a numerical solution in which optical properties can be used as a prior for computing other properties for turbid media. In this work, integrating spheres with fitted laser diodes and a monochromator was used experimentally as the irradiation source. A cryo-microtome was used to precisely slice tissue samples. Confocal and atomic force microscopy optical methods were used visualize tissue post irradiation. The results from this work offer an additional method to determine optical properties of biological tissues.

## 1.0 Introduction

Accurate determination of the optical properties of biological tissue affects the sensitivity and specificity of light based diagnostic and therapeutic tools. In the context of the integumentary system, understanding the biological impact of laser energy absorbed by skin is essential for developing models that can predict hazards from common light sources, i.e., lasers in the near-infrared range. To model and simulate possible biological damage caused by radiation in the IR spectrum, it is necessary to determine the following parameters: absorption, scattering, transmittance, anisotropy, and refractive index [1], [2]. Current research on the optical properties of tissues at laser wavelengths in the infrared (IR) spectral region of 1.0 to 4.0 μm is limited compared to the well-studied ultraviolet (UV) and visible spectra. In this work, we obtain optical properties for biological tissue at IR wavelengths in the 1-2 micrometer band and apply an analytical model to determine the optical properties based on the experimental results.

### 1.1 Optical Properties of Biological Tissue

Skin tissue is stratified and acts as a highly scattering medium for radiation in the visible and infrared wavelengths. Most of the light scattering in skin tissue is due to variations of the index of refraction or mismatched boundaries [3], [4]. Due to the intricate nature of the skin, it is a challenging problem to determine the major contributors to scattering [3]–[5]. Structurally, collagen is a major constituent of skin [4] and collagen bundles are tightly packed and less abundant in the reticular dermis, as compared to the other skin layers. Scattering in biological tissues is mostly elastic, that is, the frequency of scattered light is the same as that of the incident light and occurs due to the inherent heterogeneity of media [6], [7]. Collagen fibers have been determined to undergo Mie scattering in the IR, whereas smaller (less than 2.8 µm diameter) collagen fibers undergo Rayleigh scattering[6], [8], [9]. The Henyey-Greenstein function is commonly used to represent the influence of anisotropy, or angular dependence, in light scattering [6], [9] as shown:

$$p(\theta) = \frac{1}{4\pi} \frac{(1-g^2)}{(1+g^2-2g\cos(\theta))^{\frac{3}{2}}} \quad (1\text{-}1)$$

where "g" is the mean cosine of the scattering angle, θ, between directions of the incident and scattered photon. For biological tissue (which is forward scattering) the tissue volume has randomly distributed scatterers and thus lacks spatially correlated structures [1]. Thus, forward scattering can be represented by the first two moments of the phase function as:

$$p(\theta) = \frac{1}{4\pi} \cdot [1 + 3g(\cos(\theta))] \quad (1\text{-}2)$$

If a particle is small compared to the wavelength of incident light, there is not much variation in scattering due to direction. This is since the wavelets are in phase with each other. The phase relations for the scattered wavelets depend on scattering direction, size of the particle, and shape of the particle [6]. Consequently, as particle size increases, the phase relations tend to become distorted, and this causes a disruption in the scattering pattern. To better understand light scattering in biological specimens, it is important to consider multiple scattering events. Multiple scattering, as opposed to single scattering, refers to successive scattering of radiation within the medium.

For any given medium, scattering is quantified by the scattering coefficient, $\mu_s$. The probability of transmission, T(z), of a photon through a tissue section of thickness "z" without redirection is given as:

$$T(z) = I_o e^{\mu_a \cdot z}. \quad (1\text{-}3)$$

Monte Carlo simulations depend on the values of g and $\mu_s$ to predict the trajectory of millions of photons. Absorption is quantified through the absorption coefficient $\mu_a$ and when absorption is much greater than the scattering in a medium, the medium is a pure absorber. In this medium, light is attenuated by absorption and follows the Beer-Lambert Law (BLL) given as:

$$I(z) = I_o e^{\mu_a \cdot z} \quad (1\text{-}4)$$

where "I" is the radiance at z, $I_o$ is the incident radiation, $\mu_a$ is the absorption coefficient, and z is the path length the radiation travels within the medium [6], [10], [11].

The natural absorbers present in tissue are water and tissue chromophores. The natural chromophores present include the biological pigments; specifically, the heme pigment of hemoglobin, myoglobin, and bilirubin [1], [10], [12]. The BLL provides a macroscopic description of light absorption in media, where the intensity of the radiation decreases exponentially along the path of propagation. This model describes energy absorption by particles but does not account for scattering or emission by particles. However, radiative transfer theory can be used to account for such a medium where light scattering and absorption are both present. By knowing the absorption and scattering coefficients along with the single scattering phase function or anisotropy factor ($\mu_a$, $\mu_s$ and g) of a medium for a given wavelength, it is possible to describe the behavior of light in scattering media such as biological tissue [12], [13]. There are two types of experimental techniques employed to measure these microscopic properties: indirect techniques, and direct techniques. Optical property determination via direct techniques requires that the tissue sample be optically thin (i.e., single scattering events). Typically, this requires that the tissue sample thickness be less than the mean free path of an IR photon in the tissue. For soft tissues, this mandates a sample thickness on the order of a micron or less. Direct techniques do not depend on any specific model to obtain optical properties [6], [12]. However, while optical property determination by the direct method is simple in theory, it can be quite cumbersome and difficult in practice. Here, an indirect or analytical method was used for the determination of the tissue optical properties. Analytical methods to measure optical properties relate measured transmission and reflection values to the optical properties [13]–[15].

**1.2 Modeling**

Experimentally determined reflectance and transmittance values can be used to obtain the optical properties (i.e., $\mu_a$, $\mu_s$ and g) of biological tissue, the analysis of which is based on radiative transport theory. The fundamental basis of the photon transport theory is related by the radiative transport equation [1], [11], [16]:

$$\frac{dI(\vec{r},\hat{s})}{ds} = -\mu_t I(\vec{r},\hat{s}) + \frac{\mu_s}{4\pi}\int_{4\pi} p(\hat{s},\hat{s}') I(\vec{r},\hat{s}')d\Omega', \quad (1\text{-}5)$$

where $I(\vec{r},\hat{s})\left[\frac{W}{m^2 sr}\right]$ is the radiation density at a point r in the direction s, $p(\hat{s},\hat{s}')$ is the scattering phase function, which describes the directional distribution of scattered photons. dΩ' is the unit solid angle in the direction and µ's is the attenuation coefficient given by $\mu_t = \mu_{s+} \mu_a$, where $\mu_s$ and $\mu_a$ are the scattering and absorption coefficients, respectively. Many approaches for resolving photon transport theory are derived from this fundamental equation. Derivative methods include Monte Carlo simulations [11], [17] and Inverse Adding Doubling (IAD) [12], [13], [18], of which the IAD method was utilized here.

**1.2.1 Inverse Adding Doubling Method (IAD)**

The Inverse Adding Doubling Method (IAD) is based on the so-called "forward solution" to the transport equation using the Adding Doubling Method (ADM) [19], [20]. The ADM is a numerical solution to the transport equation whereby reflectance and transmittance values for turbid media can be calculated based on the optical properties. The IAD is the "inverse solution" which iteratively generates improved values for the albedo (a), optical thickness (τ) and anisotropy (g) from which the absorption and scattering coefficients are computed using the ADM and then compares the resulting reflectance and transmittance values [11], [13], [19]–[21]. The primary advantages of IAD over other methods used to determine optical properties are accuracy and decreased computation time in modeling mismatched boundary conditions and anisotropic scattering events [16].

The doubling method was first introduced by van de Hulst in order to solve the radiative transfer equation for a specific geometry [22]. The doubling method operates on the following assumptions: the distribution of light is independent of time, samples have homogenous optical properties, sample geometry is an infinite plane parallel slab of finite thickness, the tissue has a uniform index of refraction, internal reflections at the boundaries are governed by Fresnel's law, and the light incident upon the sample is unpolarized [14]. The advantages afforded by the ADM are: physical interpretation of results can occur at each step, and the method can be used for isotropic and anisotropic scattering. The disadvantages of ADM are in the efficiency of calculating the internal tissue fluences, that uniform irradiation must encounter the sample, and each layer is assumed to have homogenous optical properties, thereby restricting sample geometry.

### 1.3 Thickness Measurements

To use the IAD method, it is necessary to accurately measure the thickness of the tissue. To illustrate the sensitivity of the optical properties to the tissue thickness, the IAD method was used with constant values for reflectance, transmittance and index of refraction. For thick tissue sections, the optical properties are insensitive to variations in tissue thickness. However, for thin sections (<50 microns) the thickness starts to become important and for thinner sections, there is great sensitivity to changes in tissue thickness.

### 2.0 Experimental Setups

Data collected from experimental setups were conducted in the laboratories at the Applied Research Center. The test setup is devised to measure the reflectance ($R_d$), the uncollimated transmittance ($T_d$) and the collimated transmittance ($T_c$) of the tissue sample. This setup was based on the use of laser diodes as the illumination source, two integrating spheres were used for measurement of the diffuse reflectance and transmittance, and a third detector was used to measure collimated transmittance. A thermal camera was proposed to evaluate changes in tissue temperature as the test progressed. Additional test setups were devised to measure the anisotropy factor.

### 2.1 Optical Properties Test Setup

Integrating spheres (diameter 152 mm, 33 mm aperture) spatially integrate radiant flux and act as ideal optical diffusers. The spheres used here provided uniform illumination to the detectors, thereby facilitating reflectance and transmission measurements. The collimated beam from the IR source was split to provide a reference beam. To accurately measure infrared output, an IR digital lock-in radiometry detection system was utilized (Newport Merlin Control Unit). The optical chopper modulated the radiation that entered the integrating spheres, thereby allowing the radiation to be read by the PbS detectors. Thick tissue sections were placed in between the two 11 mm diaphragms thereby providing an air-tissue-air interface. The slides were cut to fit the tissue holder and the tissue was oriented such that the IR radiation encountered air-tissue, tissue-glass, and finally glass-air interfaces. To monitor the temperature of the tissue sample, an Indigo Systems TVS-700 infrared camera was used.

### 2.1.1 Monochromator Setup

To use the monochromator in the IR region of interest, a new diffraction grating with 600 g/mm, 1.6 μm blaze (Acton Research Model I1-060-1.6) was mounted on the grating turret. The light source for the monochromator was a quartz-tungsten 100 W radiation source. After aligning optical elements, the chopper frequency was set to 9 Hz. The filter on the lock-in amplifier was set to a 0.3 second time constant. The monochromator was then set to an IR wavelength, and the collimated transmittance detector was moved slightly to ensure final alignment by seeking the maximum value from the detector. After alignment, and warm up of the equipment, the tissue sample was mounted on the tissue holder. After selecting the desired wavelength, and allowing the detector signal to stabilize, each detector was measured for a time period between 30 seconds and one minute, and the average value was recorded. This procedure was repeated for various tissue samples and wavelengths.

### 2.1.2 Laser Setup

Laser diodes typically have a high beam divergence and therefore need a collimating lens. The laser setup used a 1064 nm laser and a 1313 nm laser. These lasers had a small beam divergence, so the collimating lens was removed. The 1064 nm laser was a Coherent DPSS (Diode Pumped Solid State) Nd:YAG continuous wave Class IIIb laser. The 1313 nm laser was a diode pumped Nd:YLF, Class IIIb, continuous wave system by Crystal Laser. The beam was aligned using an IR-sensitive card with final adjustment of the collimated transmittance detector performed by maximizing the reading of this detector.

### 2.1.3 Final Setup

In this final setup, (Figure 1) the laser produces collimated IR radiation which is attenuated via a variable-aperture diaphragm and chopped before entering the first integrating sphere. The detector in the first integrating sphere measured the diffuse reflectance. The detector in the second integrating sphere measured the uncollimated (diffuse) transmittance, and the third detector measures the collimated (unscattered) transmittance. Each detector was selected in turn using a selector box. The aperture on the rear diaphragm was 9 mm.

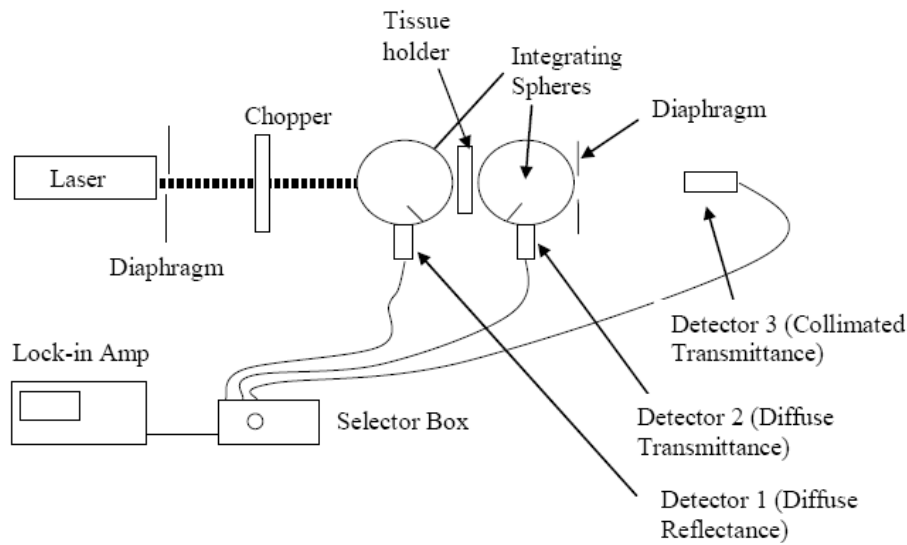

**Figure 1: Schematic Diagram of the final laser test setup.**

### 2.2 Data Acquisition and Processing

The lock-in amplifier did not have the ability to collect multiple detector data simultaneously. Therefore, the different detectors were connected to a breakout box that was sequentially switched to capture data from each detector. The data collected from the detectors were processed by the lock-in amplifier and sent to a computer via the RS232 port. A BASIC program was written to interface with the serial output of the lock-in amplifier. For each tissue sample, the following data was collected: readings from the detector in the first integrating sphere (corresponding to the diffuse reflectance or back-scatter), readings from the detector in the second integrating sphere (corresponding to the diffuse transmittance or forward-scatter), and readings from the third detector (corresponding to the collimated transmittance). A control was used to normalize the data and account for changes in the intensity of the incident beam throughout the test period. To mitigate electrical noise, room lights were kept off.

Other measurements were collected to understand the sensitivity of the setup to changes. For example, the noise in the sensors was recorded throughout the different test days to ensure that the sensors were not drifting or deteriorating. Beam intensity measurements were made with and without the tissue holder; tests were done with the tissue mounted in the rear of the first integrating sphere and in the front of the second integrating sphere, etc. To compute the optical properties from the experimental data, it was necessary to convert the raw voltage readings from the sensors into diffuse reflectance, $R_d$, diffuse transmittance, $T_d$, and collimated transmittance, $T_c$. From the measured values, the diffuse reflectance, $R_d$, is given by:

$$R_d = \frac{\frac{S_a}{I_c} - \frac{N_a}{I_{on}}}{\frac{I_{oa}}{I_{oc}} - \frac{N_a}{I_{on}}} \qquad (2\text{-}1)$$

where: $S_a$ is the signal from the sensor on the reflectance (first) integrating sphere, with the tissue sample in place. $I_c$ is the intensity of the laser beam detected from the collimated transmittance detector with no sample in place; $N_a$ is the noise detected from the sensor in the first integrating sphere when the laser is on, the sample holder is on, and there is no tissue in the sample holder; $I_{on}$ is the intensity of the laser beam detected from the collimated transmittance detector with no sample in place just after the value for $N_a$ was measured; $I_{oa}$ is the signal from the sensor on the reflectance (first) integrating sphere with no sample in place, and the exit port covered by a reflective cap, and $I_{oc}$ is the intensity of the laser beam detected from the collimated transmittance detector just after $I_{oa}$ was measured and the reflective cap was removed from the exit port of the first integrating sphere (to allow the beam through to the detector).

Similarly, the diffuse transmittance, $T_d$, is given by:

$$T_d = \frac{\frac{S_b}{I_c} - \frac{N_b}{I_{on}}}{\frac{I_{ob}}{I_{oc}} - \frac{N_b}{I_{on}}} \qquad (2\text{-}2)$$

where $S_b$ is the signal from the sensor on the transmittance (second) integrating sphere, with the tissue sample in place; $I_c$ is the intensity of the laser beam detected from the collimated transmittance detector with no tissue sample; $N_b$ is the noise detected from the sensor in the second integrating sphere when the laser is on, the sample holder is in, and there is no tissue in the sample holder; $I_{on}$ is the intensity of the laser beam detected from the collimated transmittance detector with no sample in place just after the value for $N_b$ was measured; $I_{ob}$ is the signal from the sensor on the transmittance (second) integrating sphere with no sample in place, and the exit port covered by a reflective cap and $I_{oc}$ is the intensity of the laser beam detected from the collimated transmittance detector after $I_{ob}$ was measured and the reflective cap was removed from the exit port of the second integrating sphere. Note that if the test had been performed such that the sensor readings had been obtained as a ratio between the sensor signal and the beam intensity, it would not have been necessary to divide the numbers by their respective beam intensities, I.

Finally, the collimated transmittance, $T_c$, is calculated as follows:

$$T_d = \frac{S_c}{I_c}, \qquad (2\text{-}3)$$

where $S_c$ is the signal from the collimated transmittance detector with the tissue in place, and $I_c$ is the signal from the same detector with the tissue removed. By performing the detector noise tests, the values $N_a$, $N_b$, and $I_{on}$ were obtained. By performing the beam intensity measurements, $I_{oa}$, $I_{ob}$, and $I_{oc}$ were obtained. Finally, for every tissue sample, $S_b$, $S_b$, $S_b$, and $I_c$ were collected. The calculated values for $R_d$, $T_d$, and $T_c$, were used as input values to an IAD program. This program calculated values for the albedo (a), the optical depth (τ), and the anisotropy factor, (g). From these values, and the thickness of the tissue (z), it was possible to calculate the absorption and scattering coefficients ($\mu_a$ and $\mu_s$), using the equation below:

$$\mu_s = \frac{(1-a)\cdot \tau}{z} \quad (2\text{-}4)$$

In addition to these parameters, the IAD method requires a value for the index of refraction. Tissue consists of approximately 80% water which is why the index of refraction ranges from 1.33 to about 1.60. While the index of refraction varies with wavelength, it is often approximated as 1.4 [6], [23]. This value was used for all IAD calculations in this project. To show that the results were not sensitive to the index of refraction, a fixed value was used for $R_d$, $T_d$, and $T_c$, while the index of refraction was varied between 1.3 and 1.6. The results show little change in the scattering coefficient, a 10% change in the absorption coefficient, and a 3% change in the anisotropy factor. For samples that required a glass microscope slide for structural support, the index of refraction of the slide was entered as 1.52, with an optical depth of 0.05. Early in the project, these slides were tested and shown to have high transmission in the IR.

### 3.0 Specimen Preparation

In this section, we discuss tissue preparation and extraction, followed by tissue mounting and thickness determination.

### 3.1 Tissue Preparation

The Yucatan mini pig was used as a suitable surrogate for human skin [24], [25]. All experimental protocols were approved by the Institutional Animal Care and Use Committee (IACUC). The fresh skin (18 x 17 cm) was stored and sealed in a bag with an isotonic solution or nutrient medium at a temperature between 2 and 10 degrees Celsius. Hair was systematically removed and irregular fat layers approximately 3 mm thick were removed. In addition to the full-thickness samples, thinner samples were extracted. The thinner samples were obtained using a Model B, Padgett Electro-Dermatome. The dermatome was used to cut approximately 12 0.75 mm thick strips with the epidermis and the dermis. Six of these samples included tissue with epidermis and the remaining were dermal tissue. A Tissue-Tek Microtome/Cryostat model 4551A was used to cut thin sections. This microtome cuts thin sections (2-20 micrometers) and cools down the tissue sample to -40°C. This instrument was used to cut thin samples at 10- and 20-micron thickness. Ten skin sections 1 cm x 1 cm were cut and prepped for fine cutting with the microtome at -20°C.

### 3.2 Sample Mounting

Due to the delicate nature of the tissue and the potential glass/air interaction, a custom tissue holder was constructed, as described previously. Thus, the thick samples had an air-tissue-air interface for the impinging photons. The thin samples were mounted on one slide producing air-tissue, tissue-glass, glass-air interfaces. The glass slides were cut to ensure that the tissue covered the maximum amount of the opening in the sample holder. Initial sample temperatures were confirmed using an infrared camera. Once the time required for samples to reach room temperature was determined, the infrared camera was used to provide confidence that all samples had reached room temperature prior to testing.

## 3.3 Tissue Thickness Measurement Methods

Several techniques were tested to evaluate and compare different measurement methodologies. Since the samples varied in thickness by orders of magnitude, different techniques were used for different tissue sizes. In particular, the techniques were divided into "thick" tissue measurement methods (for tissue in the millimeter range), and "thin" tissue measurements methods (for tissue in the micron range).

### 3.3.1 Thick Tissue Measurement Methods

Thick tissue measurements were obtained using three methods: an accurate hand-held micrometer, a laser profilometer, and an optical microscope with video capture capabilities.

*3.3.1.1 Micrometer Measurement Method*

One method used to measure the thickness of thicker tissue (on the order of millimeters) was to use a high-accuracy micrometer. The micrometer used for the test was a Mitutoyo model 103-135, 0-1" friction micrometer. This micrometer had a resolution of 1.0e-4" (2.5 microns) with an accuracy of ± 1e-3" and was calibrated prior to use. The diameter of the measuring faces was 0.25", with a flatness of 2.4e-4" (0.6 microns) and a parallelism of 9.0e-5" (2.3 microns). This micrometer had a friction thimble that acted as a clutch and slipped when a factory set force between the measuring surfaces was reached.

The method for measuring the thickness of the tissue was as follows. First, both slides were placed in contact with each other and the thickness of the two slides together was measured. Then, the thick samples were measured by placing the tissue between the two microscope slides and measuring the thickness of the entire assembly. Since the compression of the tissue resulted in a lower reading (i.e., thinner tissue reading), a second technique was utilized. In this method, the micrometer was adjusted until there was contact between the measuring surfaces and the slides. To ensure that there was full contact of the measuring surfaces, the micrometer was pulled in a perpendicular direction to gage the sliding resistance of the micrometer relative to the slides. The micrometer was adjusted until a slight resistance was felt when moving the micrometer (i.e., "magnetic" drag.)

*3.3.1.2 Laser Profilometer Measurement Method*

A different method utilized to measure the thickness of the tissue was the laser profilometer method. The system used for testing was the LTC model LP-2000 laser profilometer with the LP-S-5/2 sensor. This sensor works with a standoff distance between 50 and 127 mm and has a resolution of 25 microns. The sensor utilizes a red laser as the light source. The sensor was calibrated using the manufacturer's calibration block. Since this block spans the standoff distance of the sensor, a reference measurement was later made using an object approximately 7.5mm thick (measured with the micrometer) to provide a more precise calibration. The samples were placed on a slide, or a background surface of similar color to the tissue. The sample was moved such that the background distance was measured for four seconds, then the tissue was moved such that the laser targeted the center of the tissue. This position was held for four seconds and the values were recorded.

### 3.3.2 Thin Tissue Measurement Methods

Several methods were considered to measure the thickness of thin tissue. Due to the thinness of the tissue, typical optical microscopy was not suitable, nor was direct measurement using a micrometer. Possible methods considered were scanning electron microscopy (SEM), confocal microscopy and atomic force microscopy. The SEM method was not deemed to be suitable as its use would lead to specimen dehydration. However, confocal microscopy and atomic force microscopy work at atmospheric pressure and temperatures, so these methods were attempted as described below.

*3.3.2.1 Confocal Microscope*

One method used to measure the thickness of the thin specimens (on the order of microns) was to use a confocal microscope to differentially focus through the sample. The microscope used was a Leica ICM 1000 Industrial Confocal Microscope. The system specifications indicate that this system can achieve a depth (z-resolution) of better than 0.5 microns (FWHM at 635 nm wavelength). Using the microscope, it was possible to image the tissue via direct viewing through the eyepieces, or through computer imaging via photomultiplier tubes (PMTs). In order to determine the best method to measure the thickness of the tissue, different techniques were attempted. These included: acquisition of a spatial image stack using the PMTs, differential focusing through the tissue, and differential focusing from a mark on the slide to the tissue. In addition, some of these methods were repeated at two magnifications, and with and without the PMTs. One final method involved labeling the tissue with a phosphate solution. The results of these techniques are discussed in Section 5.

*3.3.2.2 Atomic Force Microscope (AFM) Method*

An atomic force microscope (AFM) (Molecular Imaging, Inc., model PicoSPM) was used to image the surface of the tissue. The sample was mounted such that the edge of the tissue was near the probe. The probe was scanned in an attempt to detect the edge of the tissue and determine the height of the probe, as it transitioned from touching the glass slide surface to the tissue surface.

**4.0 Results**

In this section the results obtained from testing are discussed. Where results are listed for multiple tissue samples under the same test conditions, a mean value was calculated along with a standard error to derive a single number that was descriptive of the results for several tissue samples.

**4.1 Monochromator**

To improve the illumination, a 250 W infrared bulb and a 600 W quartz-tungsten halogen bulb was tested. The 250W bulb also produced values at or near the noise range of the sensor. The 600 W bulb, while producing more energy, was not able to produce significantly improved results. This bulb had a much larger filament and resulted in an image that was larger than the tissue size. Therefore, the additional energy was removed by diaphragms and was not useful in increasing the signal strength. In fact, results were comparable with the 600 W bulb and the 100 W quartz-tungsten bulb, which had a dense filament. For comparison, note that the filament of the 100 W bulb measured 4.2 x 2.3 mm, whereas the filament for the 600 W lamp measured 4.0 x 13.5 mm. Moreover, with the monochromator setup, there are less safety concerns when compared to lasers. The main problem with this method was obtaining sufficient throughput from the monochromator illumination source. A possible way to overcome this limitation in the future would be to use appropriate condenser lenses transparent to NIR that would focus the filament to a smaller spot size.

**4.2 Lasers**

Using lasers, it was possible to collect useful data for the different tissue samples. The samples were numerical labeled from 1 to 34. Descriptions of the samples are found in Table 1.

**Table 1: Sample descriptions.**

| Set # | Sample Numbers | Description of Skin Samples |
|---|---|---|
| 1 | 1-11 | Approximately 20-micron thick skin |
| 2 | 12-22 | Approximately 10-micron thick skin |
| 3 | 23-28 | Approximately 0.75 mm thick skin with outer skin layer (dark) |

From these sets, some samples were rejected for being too small relative to the tissue holder aperture, resulting in a total of 30 samples tested. Due to the number of samples/laser test combinations, testing was performed over a period of several days.

**4.2.1 Laser (1064 nm wavelength)**

Using the 1064 nm wavelength laser, noise and experimental data were collected for all three sensors, along with beam intensity measurements. The raw data from these measurements were processed to provide two values for each data file (the mean value and the standard error). Using this data and the equations described above, the diffuse reflectance, diffuse transmittance and collimated transmittance were calculated for the various cases. In addition, the IAD method was used to calculate values for the albedo (a), the optical depth ($\tau$), and the anisotropy coefficient (g). During testing, the sample holder was 1° C warmer than the tissue due to the constant handling of the holder. Figure 2 below shows results for 11 optically thick and thin slices (20 micron and 10 micron respectively) irradiated at 1064 nm.

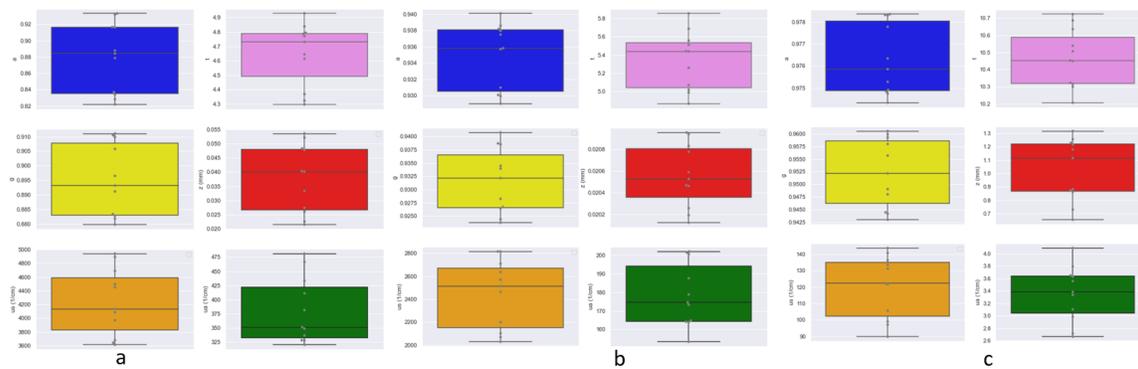

**Figure 2. Experimentally determined parameters from 20-micron thick (a), 10-micron thin (b) and 10-micron thick (c) samples with 1064 nm laser light.**

To compare the test results with the literature, results for the 1064 nm laser along, with published data are summarized in Table 2.

**Table 2: Comparison of 1064 nm results.**

| $\mu_s$ (cm$^{-1}$) | $\mu_a$ (cm$^{-1}$) | g | Material | Source |
|---|---|---|---|---|
| N/A | 0.14-0.15 | | Water | [28] |
| 220 | 0.8 | 0.88 | Porcine dermis | [25][26] |
| 120 | 0.5-2 | 0.9 | Human dermis | [27] |
| 2200 | 210 | 0.93 | Porcine epidermis | Project, 20-micron tissue |

| 4100 | 410 | 0.91 | Porcine epidermis | Project, 10-micron tissue |
| 120 | 3.2 | 0.95 | Porcine epidermis/ dermis | Project, dark thick tissue |
| 89 | 2.0 | 0.98 | Porcine dermis | Project, light thick tissue |

Variation in the collimated transmittance can cause significant variations in the optical properties computed. To further explore the effect of sample rotation, one of the samples was mounted and rotated, while data was taken from each of the three sensors (Figure 3). Results suggest that there is little variation in the diffuse reflectance and transmittance; however, there is significant variation in the collimated transmittance. Possible explanations for this variation could be variations in the sample thickness which, if coupled with a slightly off-center sample, could cause those variations. Another possibility would be that the tissue sample was slightly small or did not fill the sample holder in a small section, and rotation would have allowed more radiation to go through. While in some cases the sample was slightly small, this latter explanation seems less likely, since the smallest samples were eliminated. The sample would have to be significantly off-center and thereby produce such results.

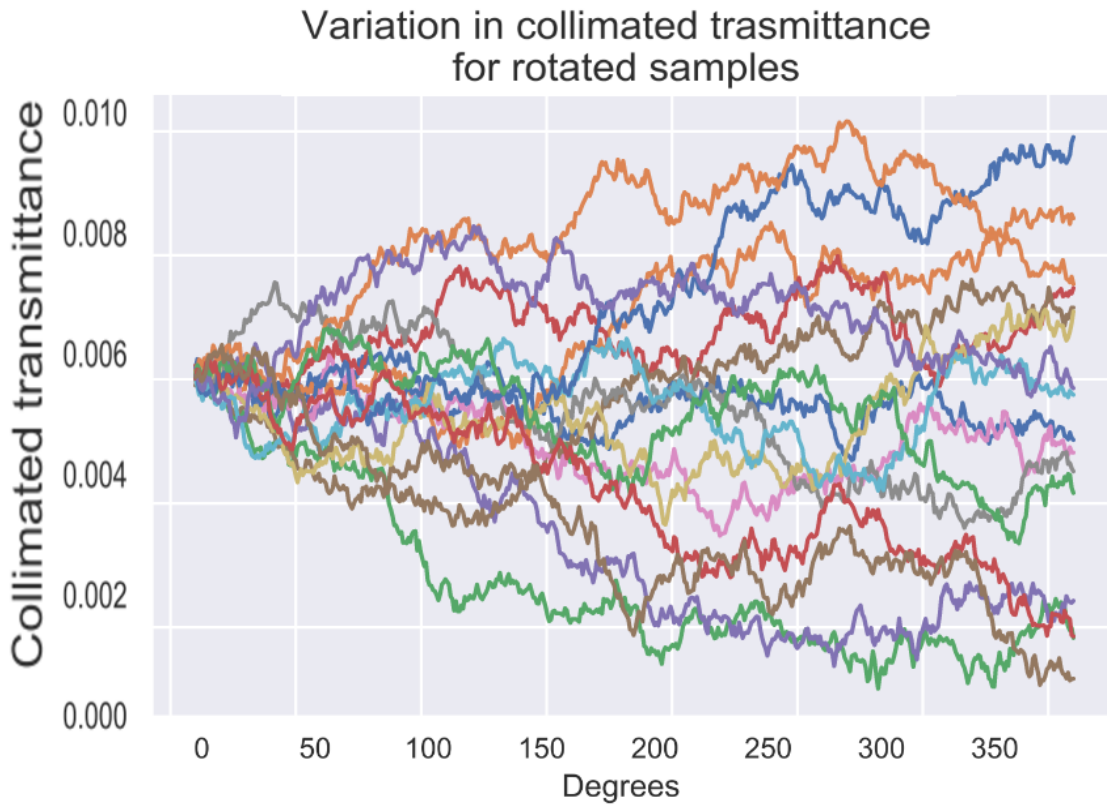

**Figure 3: Collimated transmittance sensor readings.**

**4.2.2 Laser (1310 nm Wavelength)**

The results below show computed values for the optical properties of the tissue samples based on experimental values obtained with the 1313 nm laser. Tables 3,4 shows the values for the 20-micron and 10-micron samples respectively. The resulting scattering coefficient was 2000

± 100 cm$^{-1}$ and the absorption coefficient was 180 ± 30 cm$^{-1}$. Once again, these values are nearly double the values for the 20-micron samples and cast the same uncertainty over the tissue thickness. The scattering coefficient for the tissue with epidermis was 110 ± 7 cm$^{-1}$ and the absorption coefficient was 4.6 ± 0.2 cm$^{-1}$. The scattering coefficient for the dermal tissue was 105 ± 5 cm$^{-1}$ and the absorption coefficient was 2.8 ± 0.2 cm$^{-1}$.

**Table 3: Results for 1313 nm laser, 20 micron samples.**

| Experimentally-determined values | | | | Values determined using IAD | | | | | |
|---|---|---|---|---|---|---|---|---|---|
| No. | $R_d$ | $T_d$ | $T_c$ | a | τ | g | z (mm) | $\mu_s$ (1/cm) | $\mu_a$ (1/cm) |
| S1 | 0.077 | 0.612 | 4.78E-03 | 0.966 | 5.220 | 0.963 | 0.020 | 2500 | 90 |
| S2 | 0.049 | 0.467 | 3.51E-02 | 0.860 | 3.227 | 0.943 | 0.020 | 1400 | 230 |
| S2-2 | 0.079 | 0.591 | 6.17E-03 | 0.963 | 4.965 | 0.956 | 0.020 | 2400 | 90 |
| S3 | 0.051 | 0.358 | 6.88E-03 | 0.884 | 4.856 | 0.933 | 0.020 | 2100 | 280 |
| S4 | 0.066 | 0.466 | 9.94E-03 | 0.924 | 4.488 | 0.935 | 0.020 | 2100 | 170 |
| S5 | 0.038 | 0.303 | 1.69E-02 | 0.792 | 3.955 | 0.937 | 0.020 | 1600 | 410 |
| S6-2 | 0.096 | 0.583 | 1.04E-02 | 0.965 | 4.439 | 0.937 | 0.020 | 2100 | 80 |
| S7 | 0.060 | 0.418 | 1.87E-02 | 0.890 | 3.854 | 0.916 | 0.020 | 1700 | 210 |
| S9 | 0.068 | 0.433 | 7.84E-03 | 0.922 | 4.725 | 0.928 | 0.020 | 2200 | 180 |
| S10 | 0.084 | 0.503 | 2.40E-02 | 0.932 | 3.607 | 0.907 | 0.020 | 1700 | 120 |
| S10-2 | 0.069 | 0.426 | 1.40E-02 | 0.911 | 4.143 | 0.912 | 0.020 | 1900 | 180 |
| S11 | 0.080 | 0.475 | 8.36E-03 | 0.940 | 4.661 | 0.926 | 0.020 | 2200 | 140 |
| Mean | 0.068 | 0.47 | .014 | 0.91 | 4.3 | 0.933 | 0.020 | 2000 | 180 |

**Table 4: Results for 1313 nm laser, 10 micron samples.**

| Experimentally determined values | | | | Values determined using IAD | | | | | |
|---|---|---|---|---|---|---|---|---|---|
| No. | $R_d$ | $T_d$ | $T_c$ | a | τ | g | z (mm) | $\mu_s$ (1/cm) | $\mu_a$ (1/cm) |
| S12 | 0.083 | 0.520 | 1.52E-02 | 0.943 | 4.065 | 0.924 | 0.010 | 3800 | 230 |
| S13 | 0.085 | 0.551 | 6.87E-03 | 0.959 | 4.857 | 0.943 | 0.010 | 4700 | 200 |
| S15 | 0.074 | 0.431 | 1.06E-02 | 0.922 | 4.426 | 0.914 | 0.010 | 4100 | 340 |
| S16 | 0.079 | 0.510 | 1.34E-02 | 0.940 | 4.189 | 0.928 | 0.010 | 3900 | 250 |
| S18 | 0.084 | 0.511 | 6.75E-03 | 0.951 | 4.875 | 0.935 | 0.010 | 4600 | 240 |
| S21 | 0.084 | 0.436 | 1.37E-02 | 0.927 | 4.167 | 0.899 | 0.010 | 3900 | 300 |
| S22 | 0.078 | 0.469 | 1.17E-02 | 0.933 | 4.324 | 0.920 | 0.010 | 4000 | 290 |
| Mean | 0.081 | 0.49 | 1.1E-02 | 0.939 | 4.4 | 0.923 | 0.010 | 4100 | 270 |

For purposes of comparison between test results with the literature, the results for the 1313 nm laser along, with published data are summarized in the table below.

Various sources [28]–[30] cite an absorption coefficient for water near 1313 nm at values near 1.6 cm$^{-1}$.

**Table 5: Comparison of 1313 nm results.**

| $\mu_s$ (cm$^{-1}$) | $\mu_a$ (cm$^{-1}$) | g | Material | Source |
|---|---|---|---|---|
| N/A | 1.6 | --- | Water | [28] |
| 220 | 1.2 | 0.88 | Porcine dermis | [25][26] |
| 100 | 2 | 0.9 | Human dermis | [27] |
| 2000 | 180 | 0.93 | Porcine epidermis | Project, 20 micron tissue |
| 4100 | 270 | 0.92 | Porcine epidermis | Project, 10 micron tissue |
| 110 | 4.6 | 0.89 | Porcine epidermis and dermis | Project, dark thick tissue |
| 105 | 2.8 | 0.85 | Porcine dermis | Project, light thick tissue |

**4.2.3 Age Testing**

While not a part of the scope of the project, it was decided to retest some of the tissue samples several days after it was originally tested to compare the differences in values. Table 6 has the computed results when using the 1064 nm laser, and Table 7 has the computed results when using the 1313 nm laser.

**Table 6: Retesting of thin tissue 3 days later (1064 nm laser).**

| Experimentally-determined values | | | | Values determined using IAD | | | | | |
|---|---|---|---|---|---|---|---|---|---|
| *No.* | $R_d$ | $T_d$ | $T_c$ | *a* | *τ* | *g* | *z (mm)* | $\mu_s$ (cm$^{-1}$) | $\mu_a$ (cm$^{-1}$) |
| S1 | 0.072 | 0.470 | 3.00E-03 | 0.945 | 5.686 | 0.945 | 0.020 | 2700 | 160 |
| S8 | 0.071 | 0.475 | 2.51E-03 | 0.947 | 5.863 | 0.949 | 0.020 | 2800 | 160 |
| S10 | 0.066 | 0.430 | 2.48E-02 | 0.895 | 3.573 | 0.902 | 0.020 | 1600 | 190 |
| S16 | 0.065 | 0.477 | 3.16E-02 | 0.899 | 3.331 | 0.915 | 0.010 | 3000 | 340 |
| S18 | 0.068 | 0.446 | 3.72E-03 | 0.936 | 5.470 | 0.941 | 0.010 | 5100 | 350 |
| S21 | 0.068 | 0.365 | 9.04E-03 | 0.904 | 4.583 | 0.902 | 0.010 | 4100 | 440 |
| S22 | 0.067 | 0.360 | 7.66E-03 | 0.905 | 4.748 | 0.905 | 0.010 | 4300 | 450 |

**Table 7: Retesting of thin tissue 7 days later (1313 nm laser).**

| Experimentally-determined values | | | | Values determined using IAD | | | | | |
|---|---|---|---|---|---|---|---|---|---|
| *No.* | $R_d$ | $T_d$ | $T_c$ | *a* | *τ* | *g* | *z (mm)* | $\mu_s$ (cm$^{-1}$) | $\mu_a$ (cm$^{-1}$) |
| S1-1 | 0.072 | 0.520 | 4.98E-03 | 0.949 | 5.179 | 0.950 | 0.020 | 2500 | 130 |
| S8-1 | 0.063 | 0.438 | 1.59E-02 | 0.904 | 4.017 | 0.922 | 0.020 | 1800 | 190 |
| S10-1 | 0.060 | 0.399 | 2.09E-02 | 0.881 | 3.746 | 0.906 | 0.020 | 1700 | 220 |
| S16-1 | 0.058 | 0.462 | 9.52E-03 | 0.914 | 4.531 | 0.945 | 0.010 | 4100 | 390 |
| S18-1 | 0.065 | 0.450 | 7.46E-03 | 0.925 | 4.775 | 0.936 | 0.010 | 4400 | 360 |
| S21-1 | 0.065 | 0.488 | 1.62E-02 | 0.918 | 3.999 | 0.934 | 0.010 | 3700 | 330 |
| S22-1 | 0.066 | 0.563 | 3.22E-02 | 0.924 | 3.311 | 0.940 | 0.010 | 3100 | 250 |

## 4.3 Tissue Thickness Measurement Results

As outlined in Section 3.0, several methods were used to measure the thickness of tissue samples. When significant time elapsed between testing methods, one of the earlier methods was repeated for comparison purposes.

### 4.3.1 Thick Tissue Measurement Results

The thick tissues were measured using a micrometer, an optical microscope system, and a laser profilometer. Samples labeled 23-28 were the top skin layer samples (dark samples), while samples 29-34 were the dermis samples (light samples).

*4.3.1.1 Micrometer Measurement Results*

The raw data from the thickness measurements, along with the calculated average values and standard errors are shown in Table 15. The values in Table 15 were obtained using the second method described in Section 3.3.1.1. This table shows that there is a factor of 2 variation between the minimum measured thickness and maximum measured thickness. To compare this method with the first method wherein the micrometer was used as if measuring a rigid object (i.e., using the slipping of the thimble to control the applied force at the jaws), four of the tissue samples were measured again.

*4.3.1.2 Laser Profilometer Measurement Results*

The laser profilometer was used to test the ability of this system to determine tissue thickness. In general, the laser profilometer had difficulties in reading values when surfaces of different albedos were scanned. Moreover, the system had more difficulties reading dark materials due to absorption of the laser and the resulting weakness of the reflected signal. In contrast to the difficulties in reading the dark samples, the system was able to provide measurement values for the light samples (samples 29-34). Data was collected using the laser profilometer on the light (dermis) samples. The system collected data at 40 Hz which, in four second data trials, resulted in 160 data points for each tissue measured. The first two seconds of data measured the distance to the surface on which the tissue was placed. This was the "bottom" measurement. The tissue was then moved so that the center was imaged by the profilometer for another two seconds, yielding the "top" measurement. At the moment of the switch from top to bottom measurement, there is some fluctuation in the data until it stabilized. The bottom measurement was determined by averaging samples for the first 1.5 seconds and the top measurement was determined by averaging the samples over the last 1.5 seconds (thus avoiding the measurement discontinuity at or about 2 seconds). The averaged top and bottom measurements were subtracted to obtain the tissue thickness.

Data was also collected for a reference surface, whose thickness was measured with the micrometer. The thickness of the reference surface was used to convert the voltage readings from the profilometer to distances in millimeters. For comparison, the tissue was measured using the micrometer and two glass slides as described in the previous section, since these measurements were made 23 days later. It was determined that the laser profilometer was not able to accurately determine the thickness of the light-colored tissue. Therefore, this method is not recommended for measuring the tissue thickness for any of the samples.

*4.3.1.3 Optical Microscope Measurement Results*

The tissue samples were mounted as described in Section 3.3.1.3. Samples 23-28 were imaged using the two-microscope slide method. When this method was applied to samples 29-34, there was difficulty in accurately detecting the edge of the tissue. Therefore, samples 29-34 were imaged using the single microscope slide method. Data from the measurements are shown in Table 8.

**Table 8: Optical microscope thickness measurements.**

| Sample # | Raw Measurements (mm) | | | Thickness (mm) | Std. Error ± (mm) |
|---|---|---|---|---|---|
| 23 | 0.373 | 0.419 | 0.389 | 0.394 | 0.013 |
| 24 | 0.496 | 0.494 | 0.499 | 0.496 | 0.001 |
| 25 | 0.586 | 0.675 | 0.598 | 0.620 | 0.028 |
| 26 | 0.455 | 0.463 | 0.458 | 0.459 | 0.002 |
| 27 | 0.931 | 0.777 | 0.908 | 0.872 | 0.048 |
| 28 | 0.739 | 0.647 | 0.675 | 0.687 | 0.027 |
| 29 | 1.051 | 1.015 | 1.041 | 1.036 | 0.011 |
| 30 | 0.934 | 0.862 | 0.99 | 0.929 | 0.037 |
| 31 | 0.88 | 0.872 | 0.867 | 0.873 | 0.004 |
| 32 | 1.059 | 1.023 | 1.038 | 1.040 | 0.010 |
| 33 | 0.954 | 0.967 | 0.974 | 0.965 | 0.006 |
| 34 | 0.662 | 0.637 | 0.645 | 0.648 | 0.007 |

When using this method, care must be taken, since the edge of the tissue may have slight thickness variations. When positioning the measuring lines with the software, thickness variations can be visually averaged, since the software generates a straight line that can be positioned at the average location of the edge of the tissue. To improve the measurement of the tissue thickness, various image processing methods were attempted on the dark tissue. One of these methods was a three-dimensional surface plot of pixel value (using the green plane of the RGB signal) at each x and y pixel locations. While this method could easily show the edge of the slide, the other edge of the tissue was difficult to find, since it slowly fades in brightness. Another method attempted was to use an edge-detection algorithm (Figure 4). This method was not successful in improving the results, due to the gradual change in pixel brightness.

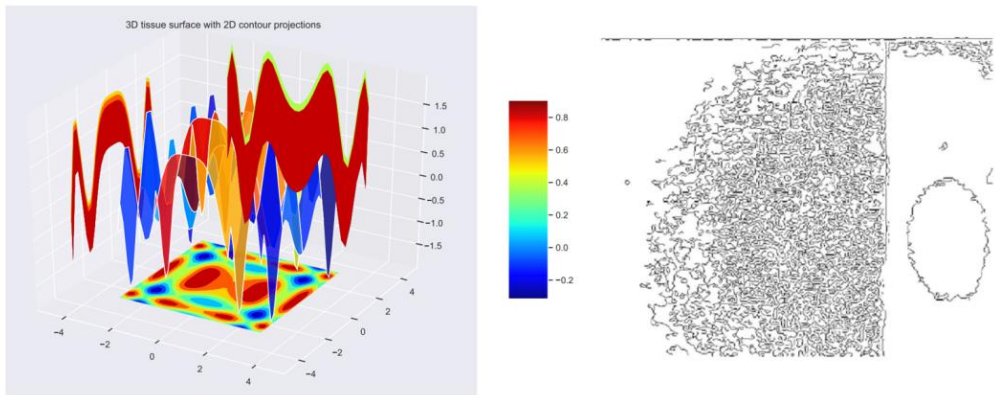

**Figure 4: Contour projections and application of an edge detection algorithm to improve measurement of tissue thickness.**

From our experiments, it was determined that the optical microscope with light tissue provides thickness measurements similar to micrometer measurements. However, the optical microscope gave poor values with the dark tissue. This was mainly due to difficulties in discerning the edges of the tissue through the microscope.

Based on these results, the best method for measuring the tissue thickness is using the micrometer on the tissue sandwiched between two slides; however, as indicated above, it is necessary to use the proper technique to not compress the tissue. Given proper illumination of the sample, the optical microscope method is also a viable option. Finally, the laser profilometer is not recommended for measuring tissue thickness.

### 4.3.2 Thin Tissue Measurement Results

*4.3.2.1 Confocal Microscope*

Using the methods described in Section 3.3.2.1, several of the samples were used in an attempt to determine their thickness using a confocal microscope. The different methods were: use of the shallow depth of field of the confocal microscope to image the top and bottom surface of the tissue by adjusting the z (vertical) position of the microscope stage, the same method repeated but using a mark on the slide to determine the bottom of the tissue, and acquisition of a spatial image stack at various depths using the photomultiplier tubes (PMTs). In addition, some of these methods were repeated at two magnifications, and with and without the PMTs. In these cases, the tissue was on a slide and no chemical additives were added. For a final test, a phosphate solution was used to image the tissue.

The first method was an attempt to view through the tissue, taking advantage of the extremely shallow depth of field designed in a confocal microscope. Attempts at visually seeing through the tissue were disappointing. It was not possible to image through the tissue in this fashion. Instead, the tissue was moved in search of anomalies such as small holes (from the animal's hair) or punctures in the tissue. Additionally, the tissue was imaged near the edge, where there would be a possibility of discerning tissue near the bottom of the sample due to irregular cutting near the edge. The idea was that such areas would allow for focusing along the side of the hole or edge of the tissue (assuming the side was visible). In some cases, it was possible to focus on the bottom of an indentation in the tissue, but it was unknown, how much distance remained between the bottom of the indentation (which still had some tissue), and the bottom of the slide. After many attempts, with low and high magnifications, the results were deemed unreliable.

Holes from other anomalies were also considered unreliable, since areas around the tears at times were thicker or thinner and not representative of the thickness of the tissue. It was also difficult to visually distinguish focusing changes for movement of the z-stage of up to 4 microns.

This method was also repeated using the PMTs to show an image of the tissue. The image was not of very high resolution and was very grainy. Moreover, it was very difficult to judge when the image was focused at a given depth. Repeated attempts were not able to produce reproducible, reasonable numbers (numbers achieved were on the order of 100 microns thickness). Other attempts were to try to image the edge of the tissue. While these areas could provide sections that could be focused, the thickness of the edge of the tissue is not necessarily representative of the thickness of the rest of the tissue. This is due to the expected thinness of the tissue due to the microtome cutting operation.

The second method attempted to make up for the difficulty in imaging the bottom surface of the tissue by making an artificial mark on the slide, adjacent to the tissue. The mark was initially made with a magic marker. Using the confocal microscope, it was difficult to determine the best focus for the mark. When the mark was made, small ink droplets were splattered nearby. The image was deemed to be in focus when the small, splattered ink droplets became visible. Even with this method, it was not possible to obtain reasonable, reproducible results. When using the PMTs, it was easy to see the ink mark in focus. However, it was difficult to determine the z-axis position of the mark, since adjustment in the z-axis did not show clear changes to the mark, unless the changes were greater than the expected thickness of the tissue. Had this method worked, there would have been a small error from the unknown thickness of the ink mark.

The third method used the PMTs and software to take images at various cross sections of the tissue and assembling them as a three-dimensional image. A limitation of this method is the z-axis movement which can only be increased at a fixed discrete step size based on the top slice z position, bottom slice z position, and the number of frames. Using this method near the edge of the tissue, it was easy to detect the bottom surface (face of the microscope slide), but it was difficult to determine the top surface of the tissue. The tissue showed many deep hills and valleys that were an order of magnitude above the expected tissue size.

Since there were difficulties in accurately determining the thickness of the tissue, several samples were processed to increase the possibility of obtaining a proper reading. These samples were labeled using two drops of a phosphate solution and covered with a cover slip. The cover slip was sealed using acetone along the periphery of the cover slip. One sample tested this way was a 5-micron sample. Measuring the thickness at different locations resulted in readings of 12.3, 11, 10.6, and 9.91 microns for an average of $11.0 \pm 0.5$. A 10-micron sample resulted in readings of 36, 25, 25.9, 27.1, and 32.4, for an average of $29 \pm 2$. A 20-micron sample was measured at four different locations producing values of 20.6, 18.8, 19.2, and 20.3 microns for an average of $19.7 \pm 0.4$ microns. These results indicate that this method worked well to measure the 20-micron sample. The discrepancies for the 5 and 10-micron sample can be due to an improper (non-reproducible) cutting technique with the microtome. These samples were taken early in the process of familiarization with the microtome. The microtome has a wheel for cutting that, if reversed, and then moved in the normal direction will result in a cut that is twice as thick. It is likely that the 5-micron sample was cut in this fashion and was actually a 10-micron sample, while the same thing may have occurred twice with the 10 micron sample producing a 30 micron sample.

*4.3.2.2 Atomic Force Microscope*

Using the methods described in Section 3.3.2.2 a tissue sample on a glass slide was placed in the atomic force microscope (AFM). One of the main limitations of the system was that it was time consuming to scan small areas of the tissue. Moreover, the scanned regions are fairly small. This necessitated trying to use the edge of the tissue to detect a step change in height. This method was not explored in full detail but rather, it was explored to determine whether it had potential for measuring the tissue thickness. Based on preliminary results, it appears that this method does have potential. While this method is a contacting method (where the AFM cantilever probe touches the sample), there was no visible adverse effect to the tissue.

**5.0 Conclusion**

Numerous different test and measurement methodologies were tried in order to find improved ways of measuring the optical properties of skin in the near infrared. In particular, three test setups were constructed to measure the reflectance, transmittance, and anisotropy factor of the tissue. Double integrating spheres equipped with a sensor in each sphere to measure diffuse reflectance and transmittance were utilized in two of the setups. In addition to the sensor, a third sensor was used to measure collimated transmittance. The main difference between the setups was the illuminating source—either a monochromator or laser was used as the source. When using the monochromator with a 100 W quartz-halogen bulb (or even a 600 W bulb) the signal values were within the noise threshold of the sensors. However, when using the lasers, it was possible to generate a signal beyond the noise level. Based on theoretical considerations, the setup for the anisotropy factor was not used, since it was believed that the samples would not be thin enough to result in single scattering of the radiation.

The measured reflectance and transmittance values were entered into an inverse adding-doubling model to calculate the optical properties of the tissue at 1064 nm and 1313 nm. The results for thin tissue were an order of magnitude above other published results; however, measurements with the thicker samples were similar to published results for human dermis. For the thick tissue, the scattering coefficient ranged between 80 and 150 $cm^{-1}$ while the absorption coefficient ranged between 1.6 and 4.1 $cm^{-1}$ when using the 1064 nm laser. For the 1313 nm laser, the scattering coefficient ranged between 81 and 141 $cm^{-1}$, while the absorption coefficient ranged between 2.5 and 5.7 $cm^{-1}$. Since the tissue samples tested were of different thickness, several methods and methods were tried to accurately measure the thickness of the tissue. These methods included micrometer readings, optical microscopy, laser profilometry, confocal microscopy, and atomic force microscopy. For thick tissue, the micrometer readings were easy to make, accurate and repeatable. However, care must be taken when using the micrometer so to not compress the tissue. Optical microscopy (with the tissue on its side) also has potential as a good method for measuring the thickness. Limitations of this method include the difficulty in determining the edges of the tissue, particularly when the sides were not cut evenly, and the depth of field limitations which may not allow focusing of both sides simultaneously.

For thin tissue, the confocal microscope and the atomic force microscope have potential for measuring the tissue thickness. With these two methods, it is necessary to image the bottom and the top of the tissue. This can be done near a tissue edge. A clean, slanted cut (relative to the depth or z- direction) should allow the AFM probe to transition from the surface of the tissue to the slide or surface where the tissue is mounted. One limitation of the confocal microscope was that it was difficult to visually discern changes in tissue focus depth when the stage is moved by as much as four microns for an unprepared sample. An unprepared sample is one where the tissue is just mounted on a slide with no chemicals added. This can result in a 40% error in measurement for a 20-micron sample (4 microns for the top measurement and 4 microns for the bottom measurement). By using a phosphate-based solution, an experienced operator was able to accurately measure the thickness of a sample. A limitation of the atomic force microscope used was that it could only scan a very small segment of the tissue. Thus, it is necessary to make sure that the edge of the tissue was properly prepared and is representative of the thickness of the rest of the tissue. Nevertheless, AFM shows potential for measuring the thickness of very thin tissue samples.

**Appendix**

**Lock-in amplifier interface to computer**

```
' MERLIN.BAS program listing:
' PROGRAM FOR DATA ACQUISITION FROM MERLIN
```

```
' AND TO SAVE DATA TO A DATA ARRAY AND TO A DISK FILE
Progloop:
CLEAR
OPTION BASE 0
INPUT "Enter output file name "; nam$
PRINT "Waiting"
z = TIMER
WHILE TIMER - z < 10
WEND
BEEP
PRINT "Sampling"
Number% = 100
OPEN "COM1:9600,N,8,1" FOR RANDOM AS #1
OPEN nam$ + ".txt" FOR OUTPUT AS #2
DIM a AS STRING * 17
PRINT " MERLIN DATA:"
z = TIMER
minrslt = 10
WHILE TIMER - z < 30
count = count + 1
PRINT #1, "PR 0 /R"
GET #1, 2, a
PRINT #1, "TD1 3 /R"
GET #1, , a
IF MID$(a, 7, 1) = "0" THEN Sign$ = "+" ELSE Esing$ = "-"
IF MID$(a, 8, 1) = "0" THEN Esing$ = "+" ELSE Esing$ = "-"
Exp$ = Esing$ + MID$(a, 9, 2)
Man$ = Sign$ + MID$(a, 12, 1) + "." + MID$(a, 13, 3)
rslt = VAL(Man$ + "E" + Exp$)
PRINT rslt
IF rslt < minrslt THEN minrslt = rslt
IF rslt > maxrslt THEN maxrslt = rslt
totrslt = totrslt + rslt
PRINT #2, Man$ + "E" + Exp$
WEND
PRINT "max value = "; maxrslt
PRINT "Min value = "; minrslt
PRINT "Average value = "; totrslt / count
CLOSE #1
```

```
CLOSE #2
BEEP
PRINT
INPUT "Continue? (y/n) "; ans$
IF ans$ <> "n" THEN GOTO Progloop
END
```